\documentclass[%
 reprint,
superscriptaddress,
 amsmath,amssymb,
 prl
]{revtex4-2}

\usepackage{graphicx}
\usepackage{dcolumn}
\usepackage{bm}
\usepackage{xcolor} 
\usepackage{soul}

\usepackage{comment}
\usepackage{braket}
\usepackage{amsmath}
\usepackage{amsfonts}
\usepackage{notes2bib}
\usepackage{bm}
\usepackage[super]{nth}

\begin{document}


\title{Meron Spin Textures in Momentum Space}

\author{Cheng Guo}
 \affiliation{Department of Applied Physics, Stanford University, Stanford, California 94305, USA}
\author{Meng Xiao}%
 \email{phmxiao@whu.edu.cn}
\affiliation{%
 Ginzton Laboratory and Department of Electrical Engineering, Stanford University, Stanford, California 94305, USA
}
\affiliation{Key Laboratory of Artificial Micro- and Nano-structures of Ministry of Education and School of Physics and Technology, Wuhan University, Wuhan 430072, China}

\author{Yu Guo}
\affiliation{%
 Ginzton Laboratory and Department of Electrical Engineering, Stanford University, Stanford, California 94305, USA
}

\author{Luqi Yuan}
\affiliation{%
 Ginzton Laboratory and Department of Electrical Engineering, Stanford University, Stanford, California 94305, USA
}
\affiliation{%
 School of Physics and Astronomy, Shanghai Jiao Tong University, Shanghai 200240, China
}

\author{Shanhui Fan}
\email{shanhui@stanford.edu}
\affiliation{%
 Ginzton Laboratory and Department of Electrical Engineering, Stanford University, Stanford, California 94305, USA
}


\date{\today}

\begin{abstract}
We reveal the meron and antimeron spin textures in momentum space in a photonic crystal slab. These spin textures in momentum space have not been previously noted either in electronic or photonic systems.  Breaking the inversion symmetry of a honeycomb photonic crystal gaps out the Dirac cones at the corners of Brillouin zone.  The spin textures of photonic bands near the gaps exhibit a meron or antimeron. Unlike the electronic systems, the spin texture of the photonic modes manifests directly in the polarization of the leakage radiation, as the Dirac points can be above the light line. The spin texture provides a direct approach to  visualize the local Berry curvature.  Our work highlights the significant opportunities of using photonic structures for the exploration of topological spin textures, with potential applications towards topologically robust ways to manipulate polarizations and other modal characteristics of light.  
\end{abstract}

\maketitle

Spin textures, the spin configuration in either real or momentum space, are of great interest in several subfields of physics. Skyrmion-related objects, including skyrmions, anti-skyrmions, merons, and anti-merons are topologically nontrivial spin textures. These textures have been extensively studied in various atomic and electronic systems such as Quantum Hall 2D electron gas, Bose-Einstein condensates, nematic liquid crystals and chiral magnets \cite{AlKhawaja2001,Roßler2006,Barrett1995,Muhlbauer2009,Yu2010, Fukuda2011, Nych2017}. Antiskymions were discovered in tetragonal Heusler materials \cite{Nayak2017}, while merons and antimerons in real space were discovered in chiral magnet thin film \cite{Yu2018}. 

\begin{figure}[htbp]
\includegraphics[width=\columnwidth]{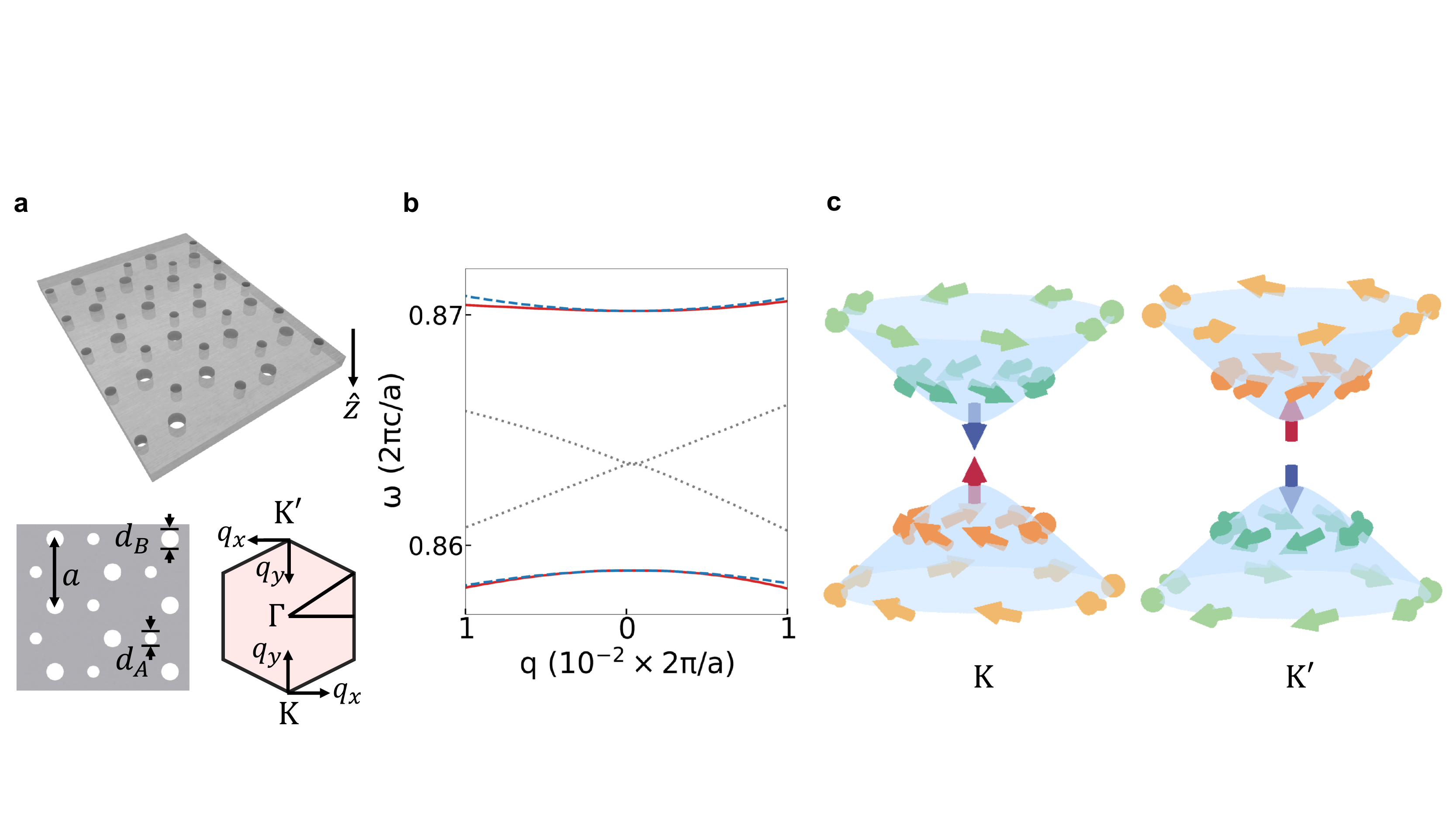}
\caption{\label{fig:Fig1} (a) A photonic crystal slab with a honeycomb lattice of circular air holes. The dielectric constant of the slab $\epsilon = 4$. The thickness of the slab is $d=0.25 a$, where $a$ is the lattice constant. The lower right shows the Brillouin zone. The wavevector  $\bm{q}=(q_x,q_y)$ measured from $K$ and $K'$ are defined individually so that $q_y$ axis points towards $\Gamma$.  (b)  The band structure near $K (K')$. The two bands form a Dirac cone when $d_A = d_B = 0.22 a$ (black dotted lines), while the degeneracy is lifted when $d_A = 0.18 a, d_B = 0.26 a$ (red). The blue dashed lines plot the fit from the effective Hamiltonian. (c) Pseudo-spin textures: core-up (down) meron for the lower (upper) band near $K$, and core-down (up) meron for the lower (upper) band near $K'$.}
\end{figure}

Since photons are massless spin-1 particles, skyrmion-related objects can also emerge as spin textures of photons \cite{VanMechelen2018,Tsesses}. Real space skyrmions have been observed recently in surface plasmon polariton systems \cite{Tsesses}. But there has not been any report of anti-skyrmions, merons and antimerons in optics. In this letter, using the honeycomb photonic crystal slab structure as shown in Fig.~\ref{fig:Fig1}(a), we report meron and antimeron in momentum space. The existence of such objects has not been previously noted either in electronic or photonic systems. The observation of such spin textures may point to topologically robust ways to manipulate polarizations of light. 

Skyrmion-related objects correspond to topologically nontrivial configurations of a three-component unit vector field $\mathbf{n}=n_x \hat{x} + n_y \hat{y} + n_z \hat{z}$ distributed over a disk in a two-dimensional space with coordinates $(x,y)$ \cite{Fert2017,Nagaosa2013}. They are all characterized by the topological skyrmion number 
\begin{equation}\label{eq:skyrmion}
    Q = \frac{1}{4\pi}\int{\mathbf{n}\cdot(\partial_x \mathbf{n} \times \partial_y \mathbf{n}) \mathop{}\!\mathrm{d}x \mathop{}\!\mathrm{d}y} , 
\end{equation}
The unit vector fields form a $2$-sphere $S^2$. For skyrmions and antiskyrmions, one considers configurations where $\mathbf{n}=\hat{z}$ at the center of the disk, and $\mathbf{n}=-\hat{z}$ at its edge. (This is referred to as the ``core-up" configuration.) Since the fields $\mathbf{n}$ are the same at the edge, one can compactify the edge to a single point to form a sphere. These field configurations thus correspond to maps of $S^2\rightarrow S^2$, which are characterized by the second homotopy group of the sphere $\pi_2(S^2) = \mathbb{Z}$, with an integer topological number $Q$ characterizing topologically distinct ways that the unit vectors wrap around the sphere. $Q=+1$ and $-1$ for skyrmions and antiskyrmions, respectively, for core-up configurations as discussed above. For core-down configurations, the signs are flipped,  i.e. $Q=-1$ and $+1$ for skyrmions and antiskyrmions, respectively.

For merons and antimerons, one considers configurations where $\mathbf{n}=\hat{z}$ at the disk center, $\mathbf{n}\bot\hat{z}$ at its edge, and $n_z\geq0$ over the whole disk. These field configurations correspond to maps of the disk to the upper  hemisphere, with the disk edge imaged to the equator. With the following map:
\begin{align}
 \label{eq:mapping}
    \mathbf{n} &= (n_x, n_y, n_z) = (\sin\theta\cos\phi, \sin \theta \sin \phi, \cos \theta )\rightarrow  \notag \\ 
    \mathbf{m} &=  (2n_xn_z, 2n_yn_z, 2n_z^2-1) = (\sin 2\theta \cos \phi , \sin 2\theta \sin \phi, \cos 2\theta )
\end{align}
which maps a hemisphere to a sphere with $0\leq\theta\leq\pi/2,  0\leq \phi \leq 2\pi$, all the points on the equator of the hemisphere are mapped to the south pole of the sphere. Applying this map to the meron or anti-meron configuration results in a field configuration with $\bm{m}=-\hat{z}$ on the edge of the disk. One can then repeat the same compactification process as the skyrmion case, and obtain an integer $Q_m$ as the topological number for $\mathbf{m}$. Since the continuous map from the  $\mathbf{n}$ field to the  $\mathbf{m}$ field doubles the solid angle subtended,  we have $Q_m = 2Q$. Therefore, merons and antimerons are characterized by half-integer skyrmion numbers: 
$Q=+1/2$ and $-1/2$ for core-up merons and  antimerons, respectively; the signs are flipped for core-down merons and antimerons \footnote{This convention of meron and antimeron is the same as most of the papers on the subject except Ref. \cite{Yu2018}, where a different convention is used.}. 

In addition to the topological number $Q$, skyrmion-related objects are further characterized by their polarity $p$ and vorticity $w$. $p=1$ for $\mathbf{n}=\hat{z}$ and $p=-1$ for $\mathbf{n}=-\hat{z}$ at the center \cite{Kovalev2018}. The vorticity $w$ indicates the rotation direction of the in-plane components of $\mathbf{n}$. Along a counterclockwise  loop around the center, for a given $w$, the in-plane components rotate an angle of $2\pi w$ counterclockwise. Skyrmions and merons have $w=1$; antiskyrmions and antimerons have $w=-1$.

Skyrmion-related objects can also emerge as spin textures of photons which are massless spin-$1$ particles \cite{VanMechelen2018,Tsesses}. Consider a polarization state as characterized by a $2\times2$ density matrix $\rho$, with the basis being the right and left circularly polarized states $\ket{\text{RCP}}$ and $\ket{\text{LCP}}$. The Stokes parameters are defined as 
$S_i=\operatorname{Tr}(\rho \sigma_i)$
where $\sigma_0 = I$; $\sigma_1=\sigma_x,\sigma_2=\sigma_y,\sigma_3=\sigma_z$ are the Pauli spin matrices \cite{Fano1949,Falkoff1951}. For a pure polarization state $\ket{\psi}$, 
$S_0^2 = S_1^2+S_2^2+S_3^2$, thus its polarization is completely characterized by a three-component unit vector, also denoted as $\mathbf{n}$:
\begin{equation} \label{eq:hatS}
    \mathbf{n} = (n_x, n_y, n_z) \equiv (S_1/S_0, S_2/S_0, S_3/S_0)
\end{equation} 
All $\mathbf{n}$'s form a unit $2$-sphere known as Poincar\'e sphere. The Poincar\'e sphere of massless spin-$1$ photon is identical to the Bloch sphere of spin-$\frac{1}{2}$ electron \cite{penrose-roadtoreality-2005}.

Here using photonic systems we show meron and anti-meron spin textures in momentum space. We consider a photonic crystal slab consisting of  a honeycomb lattice of circular air holes, where the holes at the two inequivalent sublattice sites are of different sizes  [Fig.~\ref{fig:Fig1}(a)]. For concreteness, the dielectric constant of the slab is $\epsilon = 4$, which approximates the dielectric constant of SiN at visible
wavelengths. 

The photonic band structure of the system exhibits a Dirac cone at $K$ and $K'$ when $d_A = d_B$ [black in Fig.~\ref{fig:Fig1}(b)].  Breaking the inversion symmetry ($d_A \neq d_B$) gaps out the Dirac cone, resulting in two valleys at $K$ and $K'$ \cite{Dong2017,Chen2017a}  [red in Fig.~\ref{fig:Fig1}(b)]. The system thus exhibits valley-contrasting physics similar to that in several two dimensional semiconductors \cite{Xiao2007,Cao2012}.

Breaking inversion symmetry induces meron pseudo-spin texture aound $K$ and $K'$. In the vicinity of $K$ and $K'$, the system is described by an effective Hamiltonian as obtained using the $\vec{k}\cdot \vec{p}$ method  \cite{Chen2017,Ma2016,Khanikaev2013}:  
\begin{equation}
\label{eq:Hamiltonian}
\hat{H}(q_x,q_y) = v_D (-q_y \hat{\tau}_x + q_x \hat{\tau}_y) \pm \Delta \hat{\tau}_z + \omega_0 \tau_0,
\end{equation}
where the plus (minus) sign corresponds to $K$ ($K'$). In this paper,   $\bm{q}=(q_x,q_y)$ measures the difference of the wavevector from $K$ or $K'$, with $\hat{q}_y$ axis pointing towards $\Gamma$, and $\hat{q}_x = \hat{z}\times \hat{q}_y$, where $\hat{z}$ is the unit vector perpendicular to the slab (Fig.~1a). $\bm{\hat{\tau}} = (\hat{\tau}_x, \hat{\tau}_y, \hat{\tau}_z)$ are the Pauli matrices of the pseudo-spin. $ \bm{\tau}(\bm{q})  \equiv \bra{\Psi(\mathbf{q})}\bm{\hat{\tau}}\ket{\Psi(\mathbf{q})} = (\tau_x(\bm{q}), \tau_y(\bm{q}), \tau_z(\bm{q}))$ defines the pseudo-spin texture with  $\ket{\Psi(\mathbf{q})}$ being an eigenstate at $\bm{q}$. The basis of $\bm{\hat{\tau}}$ is chosen such that $\ket{\tau_x= \pm1}$ correspond to the even/odd states with respect to the $q_y$ axis, and  $\ket{\tau_z=\pm 1}$ correspond to the  clockwise/anticlockwise-rotating states with respect to $\hat{z}$  \cite{Khanikaev2013, Chen2017a}. Below, we refer to the states $\ket{\tau_z = + 1}$ and $\ket{\tau_z = - 1}$ as the ``up" and ``down" pseudo-spin states, respectively.  $v_D$ is the group velocity. The term with $\Delta$ breaks inversion symmetry and induces a bandgap of size  $2|\Delta|$.

Figure \ref{fig:Fig1}(b) plots the eigenvalues  $E(\mathbf{q})$ of the Hamiltonian in  Eq.~(\ref{eq:Hamiltonian}) (blue dashed lines) with fitting parameters $v_D = 0.26 c, \Delta = -0.0056 \times2\pi c/a, \omega_0 = 0.8646 \times2\pi c/a$,  where $c$ is the speed of light in vacuum. $E(\mathbf{q})$ agrees well with the numerically determined photonic bands  near $K$ and $K'$ for the physical structure. 

Figure \ref{fig:Fig1}(c) depicts the pseudo-spin textures as obtained using Eq.~(\ref{eq:Hamiltonian}). At $K$ point ($\mathbf{q}=\mathbf{0}$), the pseudo-spin is up for the lower band and down for the upper band. Far away from $K$ point ($|\mathbf{q}|\gg |\Delta|/v_D$), the pseudo-spins lie in the equatorial plane with vorticity $w=1$. The pseudo-spin textures around $K$ are thus identified as core-up (core-down) meron for the lower (upper) band. Moreover, the in-plane pseudo-spin components $(\tau_x,\tau_y)$ are locked at right angles with wavevector $(q_x,q_y)$. $ \bm{\tau}(\bm{q})$ around $K'$ and $K$ are related: suppose a state in the lower band at $\bm{q}$ around $K$ has a pseudo-spin $(\tau_x,\tau_y, \tau_z)$, the corresponding state in the lower band at the same $\bm{q}$ around $K'$ has a pseudo-spin $(\tau_x,\tau_y, -\tau_z)$. The same mapping applies for the upper band. Therefore, the pseudo-spin textures around $K'$ are core-down (core-up) meron for the lower (upper) band. The meron pseudo-spin textures manifest the localized Berry curvature and the $\pm\pi$ Berry phase around $K$ and $K'$ \cite{Xiao2007}.

We proceed to show that the meron pseudo-spin textures, and hence the local Berry curvature of the photonic bands,  can be directly observed as the meron/antimeron spin texture of radiated photons. 
In our system, the valleys are above the light line since $\omega > 4 \pi c/ 3a$.  Consequently, unlike electronic systems, here the excited photonic modes will radiate out, and the leakage radiation carries information of the eigenmodes. Specifically, with respect to Fig.~\ref{fig:Fig1}(a), suppose light is incident from the $z<0$ side  with the propagation direction indicated by a unit vector $\hat{k}$. We define the S and P polarizations as having their electric field along the directions  $\hat{s}=\hat{z}\times\hat{k}$ and $\hat{p} = \hat{s}\times\hat{k}$, respectively, and the right/left circular polarization (RCP/LCP) as having their electric fields along the directions  $\hat{r} = \hat{p}+i\hat{s}$ and $ \hat{l} = \hat{p}-i\hat{s}$, respectively, where we adopt the convention of  $\exp(-i\omega t)$. 
The conventions of the Poincar\'e sphere are chosen so that $n_z=\pm 1$ correspond to  RCP/LCP,  and  $n_x= \pm1$  correspond to P/S polarizations. Now we consider the map between pseudo-spin $\bm{\tau}$ of the eigenmode and spin $\bm{n}$ of the radiated photons. The radiation process can be described by a linear map $\mathcal{F}:\ket{\Psi^i}\mapsto\ket{\Psi^{rad}}$, where $\ket{\Psi^i}$ are the internal states in the slab and $\ket{\Psi^{rad}}$ are the corresponding leakage radiation. $\ket{\Psi^i}$ can be expanded on the eigenbasis of $\ket{\tau_x = \pm 1}$, which correpsonds to even/odd states with respect to the $q_y$ axis. The even/odd states radiate into P/S polarized states only, i.e.~ $\ket{\tau_x =  1}\mapsto\ket{\text{P}}$, $\ket{\tau_x =  -1}\mapsto\ket{\text{S}}$, where the relative phase between $\ket{\text{P}}$ and $\ket{\text{S}}$ are fixed such that $\ket{\tau_z =  1} \mapsto \ket{\text{RCP}}, \ket{\tau_z =  -1} \mapsto \ket{\text{LCP}}$ at the transmission side; consequently, $\ket{\tau_z =  1} \mapsto \ket{\text{LCP}}, \ket{\tau_z =  -1} \mapsto \ket{\text{RCP}}$ at the reflection side. This map then induces a map between the pseudo-spin of photons in the slab and the spin of radiated photons as $\mathcal{F^*}:\bra{\Psi^i}\bm{\hat{\tau}}\ket{\Psi^i}\mapsto\bra{\Psi^{rad}}\bm{\hat{n}}\ket{\Psi^{rad}}$. For transmission, $(\tau_x,\tau_y,\tau_z)\mapsto(n_x,n_y,n_z)$; for reflection, $(\tau_x,\tau_y,\tau_z)\mapsto(n_x,-n_y,-n_z)$. As a result, the meron pseudo-spin textures around $K$ and $K'$ can be directly observed as meron spin textures at the transmission side and antimeron spin textures at the reflection side.

\begin{figure}[htbp]
\includegraphics[width=\columnwidth]{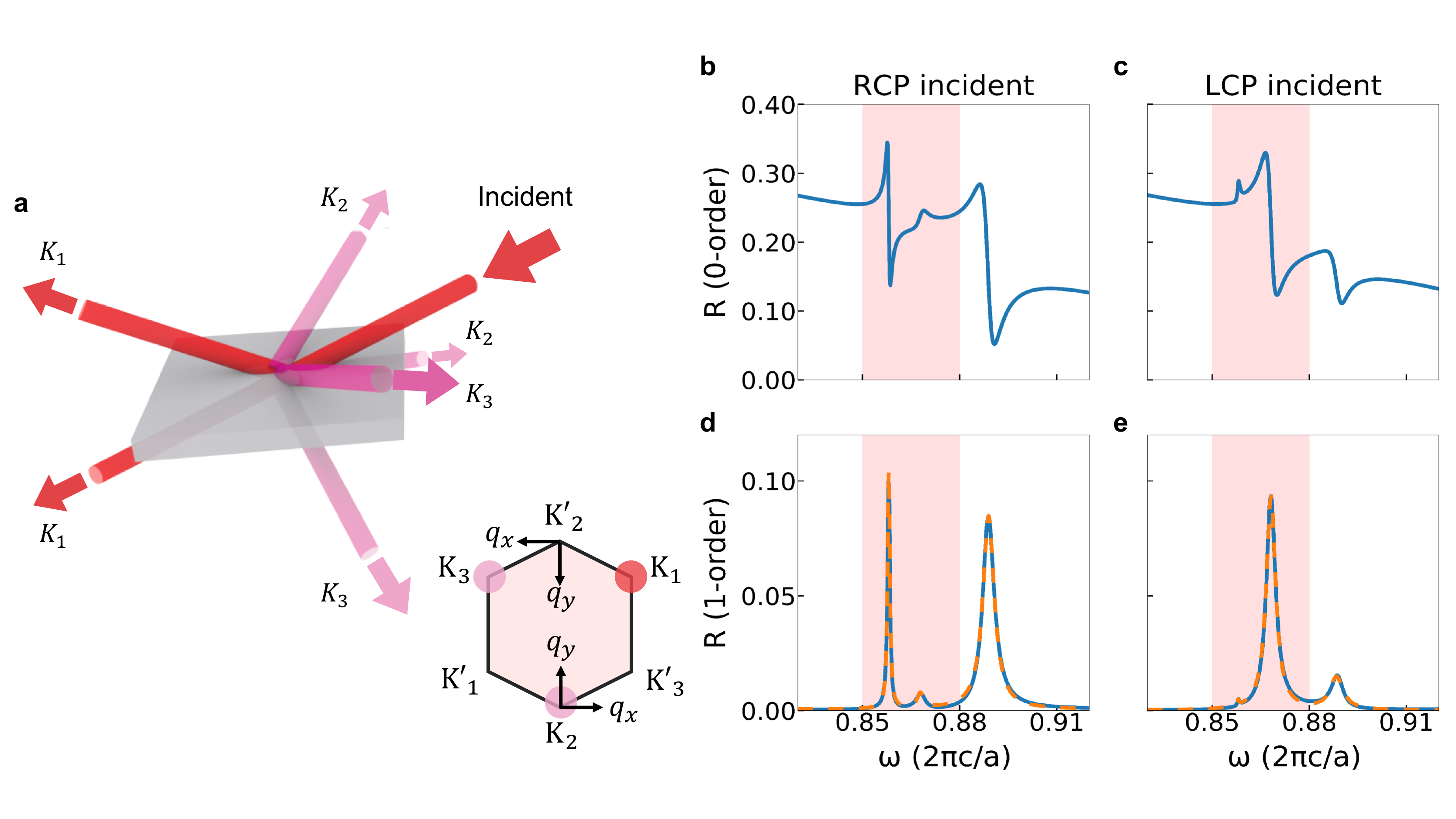}
\caption{\label{fig:Fig2} (a)  Diffraction scheme. Light with a specific frequency and polarization is incident at a specific angle to excite desired photonic modes around one Brillouin zone corner ($K_1$). The excited mode radiates out to both \nth{0}-order ($K_1$) and \nth{1}-order ($K_2$ and $K_3$) diffraction channels on both the transmission and reflection sides.  Inset: first Brillouin zone with corners $K_1,K_2$ and $K_3$ indicated.  (b-e) Calculated reflection spectra for incident light with fixed parallel wavevector $K_1$. The shaded regions include the spectral range of the photonic band gap at $K$. (b) and (c), \nth{0}-order reflection for right (b) and left (c) circularly  polarized incident light. (d) and (e), \nth{1}-order reflection for right (d) and left (e) circularly  polarized incident light. The dashed lines show the fit with Lorentzian lineshapes.}
\end{figure}

In a typical optical experiment, the modes are excited by an externally incident beam. In order to use the measured polarization properties to infer the pseudo-spin properties of the photonic modes, it is important that the light being measured contains only the radiated photons from the modes, without any interference from direct reflection/transmission of the incident bream. Therefore, we propose the setup in Fig.~\ref{fig:Fig2}(a), where we measure the polarization of light in  high-order diffraction channels.  Light with a specific frequency and polarization is incident on the sample at a specific angle to excite a desired photonic mode around one Brillouin zone corner ($K_1$). Due to the periodicity of the lattice, the excited mode radiates out to both \nth{0}-order ($K_1$) and \nth{1}-order ($K_2$ and $K_3$) channels on both the transmission and reflection sides.  Figs.~\ref{fig:Fig2}(b-e) show the calculated \nth{0}-order/\nth{1}-order reflection spectra for the RCP/LCP incident light with fixed parallel wavevector $K_1$. The \nth{0}-order spectra in Figs.~\ref{fig:Fig2}(b) and \ref{fig:Fig2}(c) exhibit Fano resonance lineshapes, superimposed upon a smoothly varying background corresponding to direct reflection \cite{Fan2002}. This indicates strong interference between the directly reflected incident light and leakage radiation from the modes in the slab. In contrast,  the \nth{1}-order spectra in Figs.~\ref{fig:Fig2}(d) and \ref{fig:Fig2}(e) exhibit resonances with Lorentzian lineshapes with negligible  background, indicating a negligible contribution from the direct reflection of the incident light. The wave amplitudes in these diffraction orders therefore arise entirely from the leakage radiation from the photonic mode in the slab. We emphasize that in this case, as long as the mode is excited, the polarization of the leakage radiation is independent of the polarization of the incident light. In general we can selectively excite either the upper or the lower band with the use of different frequencies. Near the $K$ point, where the difference in frequencies between the two bands is relatively small, we note that incident light with RCP (LCP) selectively excites the lower (upper)  state \cite{Chen2017a} at $K$ point, as shown in Fig.~\ref{fig:Fig2}(d) and \ref{fig:Fig2}(e). In this case therefore we can in addition use different polarizations of the incident light to selectively excite the upper and lower band.
\begin{figure}[htbp]
\includegraphics[width=\columnwidth]{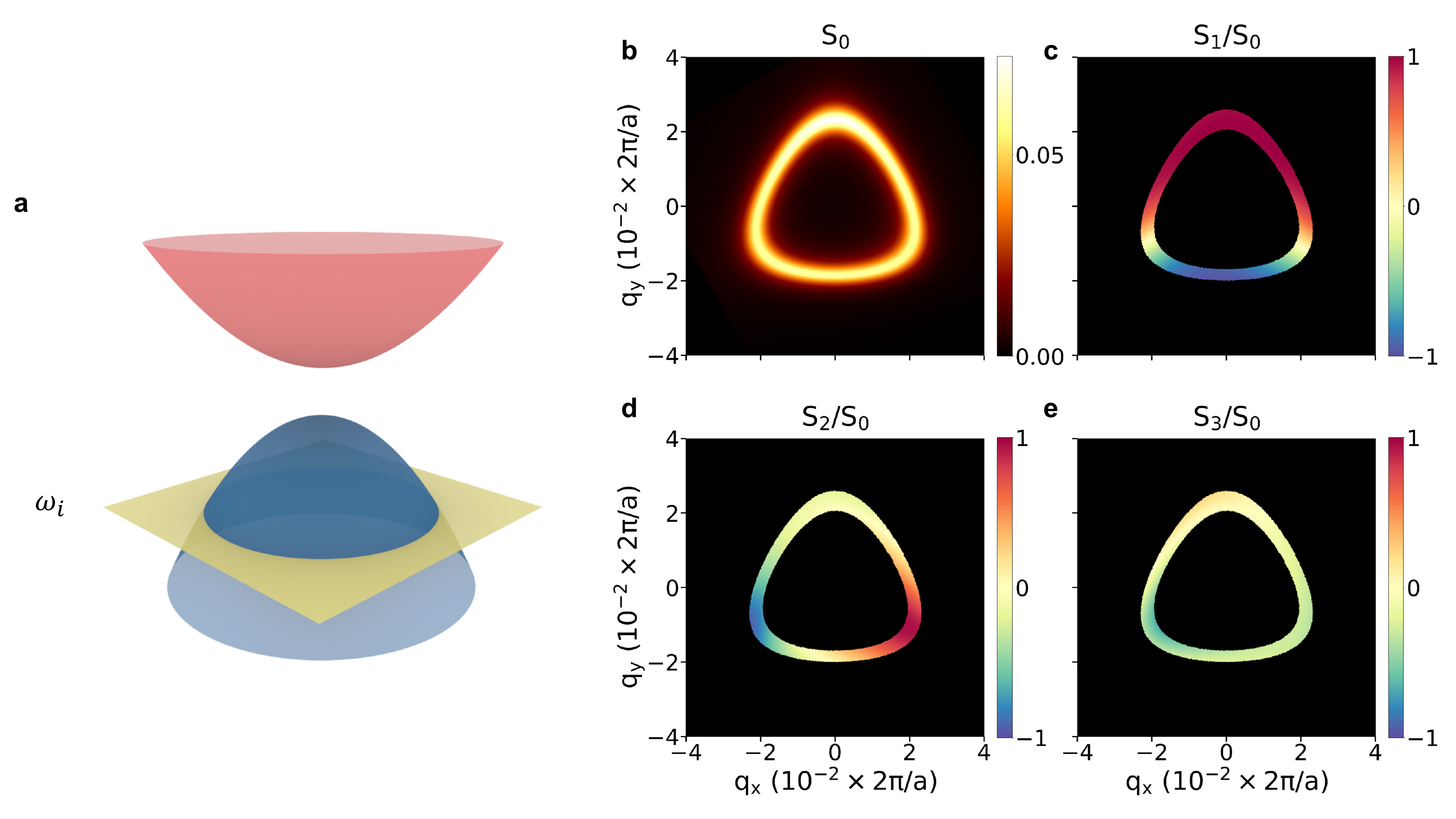}
\caption{\label{fig:Fig3} (a) Isofrequency contours of the lower band  near $K$ point are studied frequency by frequency. (b-e)  Stokes parameters as functions of $(q_x,q_y)$ at  frequency $\omega=0.855 \times 2\pi c/a$ which is in the lower band. (b) $S_0$ is the intensity of \nth{1}-order reflected light.  (c-e) Normalized Stokes parameters $S_1/S_0$, $S_2/S_0$ and $S_3/S_0$.}
\end{figure}

We now numerically study the polarization states of the photons in the \nth{1}-order  diffraction channel. The directions, frequencies and polarization of the incident light are chosen so that we probe the lower valley near $K$ point. At each frequency, we scan the incident parallel wavevectors $(k_x,k_y) = (K_{1x}+q_x, K_{1y}+q_y)$ around $K_1 = (1/\sqrt{3}, 1/3)\times 2\pi/a$, and calculate the four Stokes parameters from the electric fields of the \nth{1}-order reflected light around $K_3= (-1/\sqrt{3}, 1/3)\times 2\pi/a$ [Fig.~\ref{fig:Fig3}(a)]. Figs.~\ref{fig:Fig3}(b-e) plot the simulation results at the frequency $\omega=0.855 \times 2\pi c/a$. Figure \ref{fig:Fig3}(b) shows the intensity distribution $S_0 (q_x,q_y)$ of \nth{1}-order reflected light in momentum space, where the bright peaks match the isofrequency contour of the lower band. Figure \ref{fig:Fig3}(c-e) show normalized Stokes parameters $S_1/S_0(q_x,q_y)$, $S_2/S_0(q_x,q_y)$ and $S_3/S_0(q_x,q_y)$, respectively. Since the Stoke parameters are not well defined for $S_0 = 0$, we only show results for $S_0 > 0.04$. The polarizations  show significant variation in the direction along the isofrequency contour, but far less variation in the direction perpendicular to the contour. This is consistent with the fact that the spin texture of the leakage radiation manifests the pseudo-spin texture of the underlying photonic modes in this setup.  
\begin{figure}[htbp]
\includegraphics[width=\columnwidth]{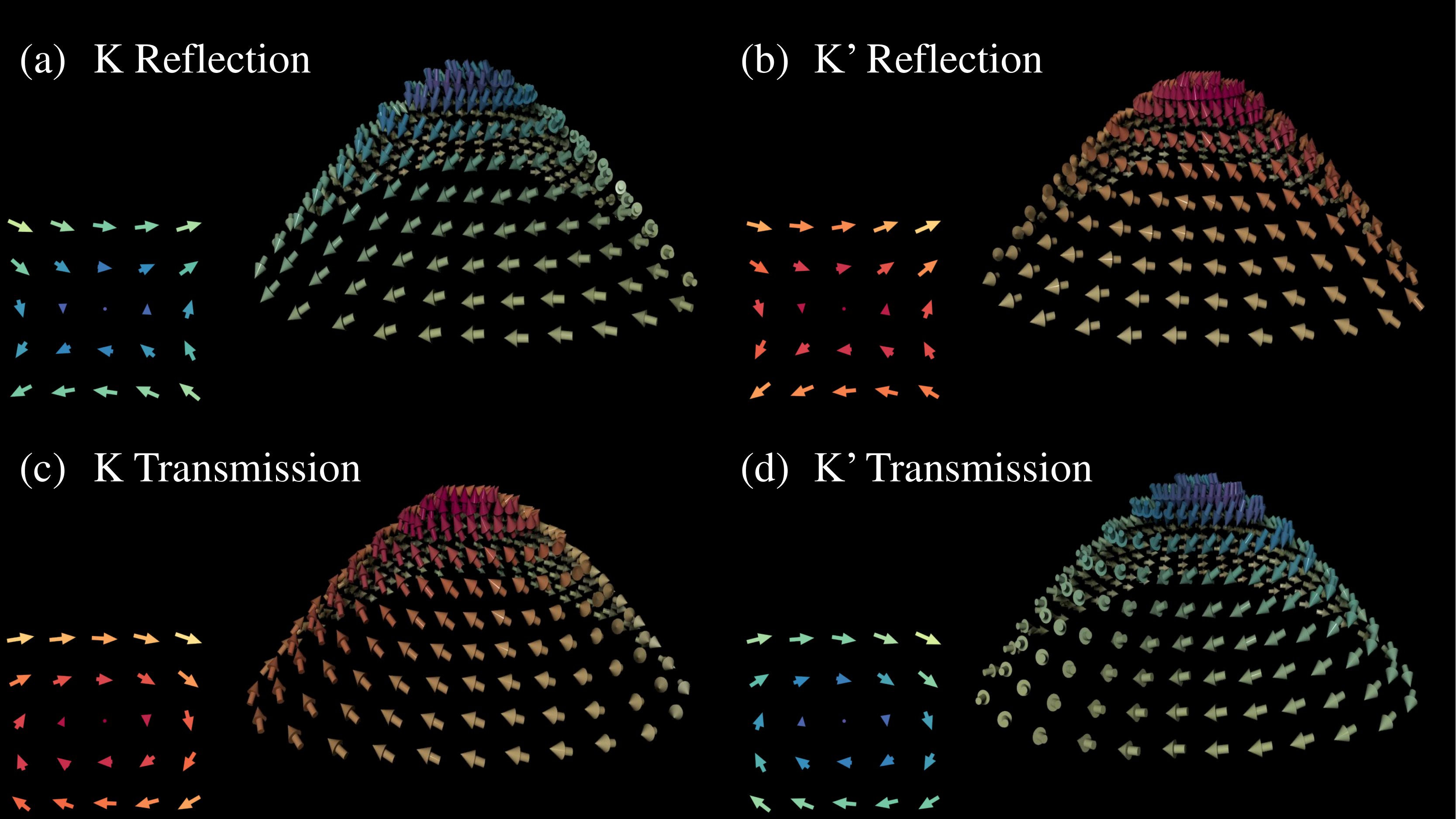}
\caption{\label{fig:Fig4} In the main figure of each panel, the arrow tail positions indicate the band dispersion $\omega(q_x, q_y)$. The arrow direction indicates the spin $\mathbf{n}$ at that point. The inset plots the in-plane spin component $(S_1/S_0,S_2/S_0)(q_x,q_y)$. (a) $K$ valley, reflection. The spin texture is a  core-down antimeron (Skyrmion number $Q=1/2$,  polarity $p = -1$, vorticity $w=-1$). (b) $K'$ valley, reflection. The spin texture is a core-up antimeron ($Q=-1/2$, $p = 1$, $w=-1$). (c) $K$ valley, transmission. The spin texture is a core-up meron ($Q=1/2$, $p = 1$, $w=1$). This spin texture is identically mapped from the pseudo-spin texture near $K$ [Fig.~\ref{fig:Fig1}(c)].  (d) $K'$ valley, transmission. The spin texture is a core-down meron ($Q=-1/2$, $p = -1$, $w=1$). This spin texture is identically mapped from the pseudo-spin texture near $K'$ [Fig.~\ref{fig:Fig1}(c)].}
\end{figure}

In Fig.~\ref{fig:Fig4}, we plot the spin textures of the leakage radiation on the iso-frequency contours of the photonic band structure. Near the $K$-valley, for the reflected light in the \nth{1}-order channel, the texture corresponds to a core-down antimeron with skymion number $Q=1/2$, polarity $p=-1$ and vorticity $w=-1$  [Fig.~\ref{fig:Fig4}(a)]. At the $K$ point, the spin points down ($\bm{n} = (0,0,-1)$), corresponding to LCP. Away from the $K$-point the spin gradually rotates to the equatorial plane, corresponding to linearly-polarized light. Notice that the in-plane spin components on the circle around the $K$-point have a winding number of $-1$.  At the $K'$-valley, the spin texture of the reflected light corresponds to a core-up anti-meron with  $Q=-1/2,p=1,w=-1$  [Fig.~\ref{fig:Fig4}(b)]. This texture has the same winding characteristics as the texture shown in Fig.~\ref{fig:Fig4}(a), but with spin up at the core of $K'$. For the transmitted light in the \nth{1}-order channel, the texture corresponds to a core-up meron at $K$ with $Q=1/2,p=1,w=1$ [Fig.~\ref{fig:Fig4}(c)], and a core-down meron at $K’$ with $Q=-1/2,p=-1,w=1$ [Fig.~\ref{fig:Fig4}(d)]. Notice the in-plane spin components have a winding number of $+1$. The relation of spin textures between the transmitted and reflected lights can be explained by the mirror symmetry of the modes in the slab, whereas the relation of the textures between the $K$ and $K'$ valley can be explained by time-reversal symmetry and the adopted coordinate system. The observed spin texture of the leakage radiation corresponds well to the pseudo-spin texture of the photonic modes in the slab as described by Eq.~(\ref{eq:Hamiltonian}). The analysis of the leakage radiation provides a direct visualization of the intriguing connection of spin, pseudo-spin, valley and band topology in the photonic valleytronic systems. In particular, our setup directly maps out the Berry curvature, which has only been probed indirectly by wave packet transport \cite{wimmer2017experimental}.  

In conclusion, we reveal the intrinsic meron pseudo-spin texture in momentum space in a photonic crystal slab, which can be directly observed as meron and antimeron spin texture by polarimetric study of high-order diffracted light from the system. Such spin texture in momentum space has not been previously observed in either electronic or photonic systems. Our work indicates significant opportunities of using photonic structures to explore topologically non-trivial spin textures. Our result may also be important for arbitrary polarization generation  \cite{Guo2019,Zhang2018,Doeleman2018}. For example, in this system, by changing the angle of incidence near the $K$ point by a small amount, a wide variety of different polarizations can be generated. 

This work is supported by the National Science Foundation (CBET-1641069).


\begin{thebibliography}{30}%
\makeatletter
\providecommand \@ifxundefined [1]{%
 \@ifx{#1\undefined}
}%
\providecommand \@ifnum [1]{%
 \ifnum #1\expandafter \@firstoftwo
 \else \expandafter \@secondoftwo
 \fi
}%
\providecommand \@ifx [1]{%
 \ifx #1\expandafter \@firstoftwo
 \else \expandafter \@secondoftwo
 \fi
}%
\providecommand \natexlab [1]{#1}%
\providecommand \enquote  [1]{``#1''}%
\providecommand \bibnamefont  [1]{#1}%
\providecommand \bibfnamefont [1]{#1}%
\providecommand \citenamefont [1]{#1}%
\providecommand \href@noop [0]{\@secondoftwo}%
\providecommand \href [0]{\begingroup \@sanitize@url \@href}%
\providecommand \@href[1]{\@@startlink{#1}\@@href}%
\providecommand \@@href[1]{\endgroup#1\@@endlink}%
\providecommand \@sanitize@url [0]{\catcode `\\12\catcode `\$12\catcode
  `\&12\catcode `\#12\catcode `\^12\catcode `\_12\catcode `\%12\relax}%
\providecommand \@@startlink[1]{}%
\providecommand \@@endlink[0]{}%
\providecommand \url  [0]{\begingroup\@sanitize@url \@url }%
\providecommand \@url [1]{\endgroup\@href {#1}{\urlprefix }}%
\providecommand \urlprefix  [0]{URL }%
\providecommand \Eprint [0]{\href }%
\providecommand \doibase [0]{https://doi.org/}%
\providecommand \selectlanguage [0]{\@gobble}%
\providecommand \bibinfo  [0]{\@secondoftwo}%
\providecommand \bibfield  [0]{\@secondoftwo}%
\providecommand \translation [1]{[#1]}%
\providecommand \BibitemOpen [0]{}%
\providecommand \bibitemStop [0]{}%
\providecommand \bibitemNoStop [0]{.\EOS\space}%
\providecommand \EOS [0]{\spacefactor3000\relax}%
\providecommand \BibitemShut  [1]{\csname bibitem#1\endcsname}%
\let\auto@bib@innerbib\@empty
\bibitem [{\citenamefont {{Al Khawaja}}\ and\ \citenamefont
  {Stoof}(2001)}]{AlKhawaja2001}%
  \BibitemOpen
  \bibfield  {author} {\bibinfo {author} {\bibfnamefont {U.}~\bibnamefont {{Al
  Khawaja}}}\ and\ \bibinfo {author} {\bibfnamefont {H.}~\bibnamefont
  {Stoof}},\ }\bibfield  {title} {\bibinfo {title} {{Skyrmions in a
  ferromagnetic Bose-Einstein condensate}},\ }\href
  {https://doi.org/10.1038/35082010} {\bibfield  {journal} {\bibinfo  {journal}
  {Nature}\ }\textbf {\bibinfo {volume} {411}},\ \bibinfo {pages} {918}
  (\bibinfo {year} {2001})}\BibitemShut {NoStop}%
\bibitem [{\citenamefont {R{\"{o}}{\ss}ler}\ \emph {et~al.}(2006)\citenamefont
  {R{\"{o}}{\ss}ler}, \citenamefont {Bogdanov},\ and\ \citenamefont
  {Pfleiderer}}]{Roßler2006}%
  \BibitemOpen
  \bibfield  {author} {\bibinfo {author} {\bibfnamefont {U.~K.}\ \bibnamefont
  {R{\"{o}}{\ss}ler}}, \bibinfo {author} {\bibfnamefont {A.~N.}\ \bibnamefont
  {Bogdanov}},\ and\ \bibinfo {author} {\bibfnamefont {C.}~\bibnamefont
  {Pfleiderer}},\ }\bibfield  {title} {\bibinfo {title} {{Spontaneous skyrmion
  ground states in magnetic metals}},\ }\href
  {https://doi.org/10.1038/nature05056} {\bibfield  {journal} {\bibinfo
  {journal} {Nature}\ }\textbf {\bibinfo {volume} {442}},\ \bibinfo {pages}
  {797} (\bibinfo {year} {2006})}\BibitemShut {NoStop}%
\bibitem [{\citenamefont {Barrett}\ \emph {et~al.}(1995)\citenamefont
  {Barrett}, \citenamefont {Dabbagh}, \citenamefont {Pfeiffer}, \citenamefont
  {West},\ and\ \citenamefont {Tycko}}]{Barrett1995}%
  \BibitemOpen
  \bibfield  {author} {\bibinfo {author} {\bibfnamefont {S.~E.}\ \bibnamefont
  {Barrett}}, \bibinfo {author} {\bibfnamefont {G.}~\bibnamefont {Dabbagh}},
  \bibinfo {author} {\bibfnamefont {L.~N.}\ \bibnamefont {Pfeiffer}}, \bibinfo
  {author} {\bibfnamefont {K.~W.}\ \bibnamefont {West}},\ and\ \bibinfo
  {author} {\bibfnamefont {R.}~\bibnamefont {Tycko}},\ }\bibfield  {title}
  {\bibinfo {title} {{Optically pumped NMR evidence for finite-size Skyrmions
  in GaAs quantum wells near Landau level filling $\nu$=1}},\ }\href
  {https://doi.org/10.1103/PhysRevLett.74.5112} {\bibfield  {journal} {\bibinfo
   {journal} {Phys. Rev. Lett.}\ }\textbf {\bibinfo {volume} {74}},\ \bibinfo
  {pages} {5112} (\bibinfo {year} {1995})}\BibitemShut {NoStop}%
\bibitem [{\citenamefont {Muhlbauer}\ \emph {et~al.}(2009)\citenamefont
  {Muhlbauer}, \citenamefont {Binz}, \citenamefont {Jonietz}, \citenamefont
  {Pfleiderer}, \citenamefont {Rosch}, \citenamefont {Neubauer}, \citenamefont
  {Georgii},\ and\ \citenamefont {Boni}}]{Muhlbauer2009}%
  \BibitemOpen
  \bibfield  {author} {\bibinfo {author} {\bibfnamefont {S.}~\bibnamefont
  {Muhlbauer}}, \bibinfo {author} {\bibfnamefont {B.}~\bibnamefont {Binz}},
  \bibinfo {author} {\bibfnamefont {F.}~\bibnamefont {Jonietz}}, \bibinfo
  {author} {\bibfnamefont {C.}~\bibnamefont {Pfleiderer}}, \bibinfo {author}
  {\bibfnamefont {A.}~\bibnamefont {Rosch}}, \bibinfo {author} {\bibfnamefont
  {A.}~\bibnamefont {Neubauer}}, \bibinfo {author} {\bibfnamefont
  {R.}~\bibnamefont {Georgii}},\ and\ \bibinfo {author} {\bibfnamefont
  {P.}~\bibnamefont {Boni}},\ }\bibfield  {title} {\bibinfo {title} {{Skyrmion
  Lattice in a Chiral Magnet}},\ }\href
  {https://doi.org/10.1126/science.1166767} {\bibfield  {journal} {\bibinfo
  {journal} {Science}\ }\textbf {\bibinfo {volume} {323}},\ \bibinfo {pages}
  {915} (\bibinfo {year} {2009})}\BibitemShut {NoStop}%
\bibitem [{\citenamefont {Yu}\ \emph {et~al.}(2010)\citenamefont {Yu},
  \citenamefont {Onose}, \citenamefont {Kanazawa}, \citenamefont {Park},
  \citenamefont {Han}, \citenamefont {Matsui}, \citenamefont {Nagaosa},\ and\
  \citenamefont {Tokura}}]{Yu2010}%
  \BibitemOpen
  \bibfield  {author} {\bibinfo {author} {\bibfnamefont {X.~Z.}\ \bibnamefont
  {Yu}}, \bibinfo {author} {\bibfnamefont {Y.}~\bibnamefont {Onose}}, \bibinfo
  {author} {\bibfnamefont {N.}~\bibnamefont {Kanazawa}}, \bibinfo {author}
  {\bibfnamefont {J.~H.}\ \bibnamefont {Park}}, \bibinfo {author}
  {\bibfnamefont {J.~H.}\ \bibnamefont {Han}}, \bibinfo {author} {\bibfnamefont
  {Y.}~\bibnamefont {Matsui}}, \bibinfo {author} {\bibfnamefont
  {N.}~\bibnamefont {Nagaosa}},\ and\ \bibinfo {author} {\bibfnamefont
  {Y.}~\bibnamefont {Tokura}},\ }\bibfield  {title} {\bibinfo {title}
  {{Real-space observation of a two-dimensional skyrmion crystal}},\ }\href
  {https://doi.org/10.1038/nature09124} {\bibfield  {journal} {\bibinfo
  {journal} {Nature}\ }\textbf {\bibinfo {volume} {465}},\ \bibinfo {pages}
  {901} (\bibinfo {year} {2010})}\BibitemShut {NoStop}%
\bibitem [{\citenamefont {Fukuda}\ and\ \citenamefont
  {{\v{Z}}umer}(2011)}]{Fukuda2011}%
  \BibitemOpen
  \bibfield  {author} {\bibinfo {author} {\bibfnamefont {J.-i.}\ \bibnamefont
  {Fukuda}}\ and\ \bibinfo {author} {\bibfnamefont {S.}~\bibnamefont
  {{\v{Z}}umer}},\ }\bibfield  {title} {\bibinfo {title}
  {{Quasi-two-dimensional Skyrmion lattices in a chiral nematic liquid
  crystal}},\ }\href {https://doi.org/10.1038/ncomms1250} {\bibfield  {journal}
  {\bibinfo  {journal} {Nat. Commun.}\ }\textbf {\bibinfo {volume} {2}},\
  \bibinfo {pages} {246} (\bibinfo {year} {2011})}\BibitemShut {NoStop}%
\bibitem [{\citenamefont {Nych}\ \emph {et~al.}(2017)\citenamefont {Nych},
  \citenamefont {Fukuda}, \citenamefont {Ognysta}, \citenamefont
  {{\v{Z}}umer},\ and\ \citenamefont {Mu{\v{s}}evi{\v{c}}}}]{Nych2017}%
  \BibitemOpen
  \bibfield  {author} {\bibinfo {author} {\bibfnamefont {A.}~\bibnamefont
  {Nych}}, \bibinfo {author} {\bibfnamefont {J.-i.}\ \bibnamefont {Fukuda}},
  \bibinfo {author} {\bibfnamefont {U.}~\bibnamefont {Ognysta}}, \bibinfo
  {author} {\bibfnamefont {S.}~\bibnamefont {{\v{Z}}umer}},\ and\ \bibinfo
  {author} {\bibfnamefont {I.}~\bibnamefont {Mu{\v{s}}evi{\v{c}}}},\ }\bibfield
   {title} {\bibinfo {title} {{Spontaneous formation and dynamics of
  half-skyrmions in a chiral liquid-crystal film}},\ }\href
  {https://doi.org/10.1038/nphys4245} {\bibfield  {journal} {\bibinfo
  {journal} {Nat. Phys.}\ }\textbf {\bibinfo {volume} {13}},\ \bibinfo {pages}
  {1215} (\bibinfo {year} {2017})}\BibitemShut {NoStop}%
\bibitem [{\citenamefont {Nayak}\ \emph {et~al.}(2017)\citenamefont {Nayak},
  \citenamefont {Kumar}, \citenamefont {Ma}, \citenamefont {Werner},
  \citenamefont {Pippel}, \citenamefont {Sahoo}, \citenamefont {Damay},
  \citenamefont {R{\"{o}}{\ss}ler}, \citenamefont {Felser},\ and\ \citenamefont
  {Parkin}}]{Nayak2017}%
  \BibitemOpen
  \bibfield  {author} {\bibinfo {author} {\bibfnamefont {A.~K.}\ \bibnamefont
  {Nayak}}, \bibinfo {author} {\bibfnamefont {V.}~\bibnamefont {Kumar}},
  \bibinfo {author} {\bibfnamefont {T.}~\bibnamefont {Ma}}, \bibinfo {author}
  {\bibfnamefont {P.}~\bibnamefont {Werner}}, \bibinfo {author} {\bibfnamefont
  {E.}~\bibnamefont {Pippel}}, \bibinfo {author} {\bibfnamefont
  {R.}~\bibnamefont {Sahoo}}, \bibinfo {author} {\bibfnamefont
  {F.}~\bibnamefont {Damay}}, \bibinfo {author} {\bibfnamefont {U.~K.}\
  \bibnamefont {R{\"{o}}{\ss}ler}}, \bibinfo {author} {\bibfnamefont
  {C.}~\bibnamefont {Felser}},\ and\ \bibinfo {author} {\bibfnamefont
  {S.~S.~P.}\ \bibnamefont {Parkin}},\ }\bibfield  {title} {\bibinfo {title}
  {{Magnetic antiskyrmions above room temperature in tetragonal Heusler
  materials}},\ }\href {https://doi.org/10.1038/nature23466} {\bibfield
  {journal} {\bibinfo  {journal} {Nature}\ }\textbf {\bibinfo {volume} {548}},\
  \bibinfo {pages} {561} (\bibinfo {year} {2017})}\BibitemShut {NoStop}%
\bibitem [{\citenamefont {Yu}\ \emph {et~al.}(2018)\citenamefont {Yu},
  \citenamefont {Koshibae}, \citenamefont {Tokunaga}, \citenamefont {Shibata},
  \citenamefont {Taguchi}, \citenamefont {Nagaosa},\ and\ \citenamefont
  {Tokura}}]{Yu2018}%
  \BibitemOpen
  \bibfield  {author} {\bibinfo {author} {\bibfnamefont {X.~Z.}\ \bibnamefont
  {Yu}}, \bibinfo {author} {\bibfnamefont {W.}~\bibnamefont {Koshibae}},
  \bibinfo {author} {\bibfnamefont {Y.}~\bibnamefont {Tokunaga}}, \bibinfo
  {author} {\bibfnamefont {K.}~\bibnamefont {Shibata}}, \bibinfo {author}
  {\bibfnamefont {Y.}~\bibnamefont {Taguchi}}, \bibinfo {author} {\bibfnamefont
  {N.}~\bibnamefont {Nagaosa}},\ and\ \bibinfo {author} {\bibfnamefont
  {Y.}~\bibnamefont {Tokura}},\ }\bibfield  {title} {\bibinfo {title}
  {{Transformation between meron and skyrmion topological spin textures in a
  chiral magnet}},\ }\href {https://doi.org/10.1038/s41586-018-0745-3}
  {\bibfield  {journal} {\bibinfo  {journal} {Nature}\ }\textbf {\bibinfo
  {volume} {564}},\ \bibinfo {pages} {95} (\bibinfo {year} {2018})}\BibitemShut
  {NoStop}%
\bibitem [{\citenamefont {Van~Mechelen}\ and\ \citenamefont
  {Jacob}(2019)}]{VanMechelen2018}%
  \BibitemOpen
  \bibfield  {author} {\bibinfo {author} {\bibfnamefont {T.}~\bibnamefont
  {Van~Mechelen}}\ and\ \bibinfo {author} {\bibfnamefont {Z.}~\bibnamefont
  {Jacob}},\ }\bibfield  {title} {\bibinfo {title} {{Photonic Dirac monopoles
  and skyrmions: spin-1 quantization}},\ }\href
  {https://doi.org/10.1364/OME.9.000095} {\bibfield  {journal} {\bibinfo
  {journal} {Opt. Mater. Express}\ }\textbf {\bibinfo {volume} {9}},\ \bibinfo
  {pages} {95} (\bibinfo {year} {2019})}\BibitemShut {NoStop}%
\bibitem [{\citenamefont {Tsesses}\ \emph {et~al.}(2018)\citenamefont
  {Tsesses}, \citenamefont {Ostrovsky}, \citenamefont {Cohen}, \citenamefont
  {Gjonaj}, \citenamefont {Lindner},\ and\ \citenamefont {Bartal}}]{Tsesses}%
  \BibitemOpen
  \bibfield  {author} {\bibinfo {author} {\bibfnamefont {S.}~\bibnamefont
  {Tsesses}}, \bibinfo {author} {\bibfnamefont {E.}~\bibnamefont {Ostrovsky}},
  \bibinfo {author} {\bibfnamefont {K.}~\bibnamefont {Cohen}}, \bibinfo
  {author} {\bibfnamefont {B.}~\bibnamefont {Gjonaj}}, \bibinfo {author}
  {\bibfnamefont {N.~H.}\ \bibnamefont {Lindner}},\ and\ \bibinfo {author}
  {\bibfnamefont {G.}~\bibnamefont {Bartal}},\ }\bibfield  {title} {\bibinfo
  {title} {{Optical skyrmion lattice in evanescent electromagnetic fields}},\
  }\href {https://doi.org/10.1126/science.aau0227} {\bibfield  {journal}
  {\bibinfo  {journal} {Science}\ }\textbf {\bibinfo {volume} {361}},\ \bibinfo
  {pages} {993} (\bibinfo {year} {2018})}\BibitemShut {NoStop}%
\bibitem [{\citenamefont {Fert}\ \emph {et~al.}(2017)\citenamefont {Fert},
  \citenamefont {Reyren},\ and\ \citenamefont {Cros}}]{Fert2017}%
  \BibitemOpen
  \bibfield  {author} {\bibinfo {author} {\bibfnamefont {A.}~\bibnamefont
  {Fert}}, \bibinfo {author} {\bibfnamefont {N.}~\bibnamefont {Reyren}},\ and\
  \bibinfo {author} {\bibfnamefont {V.}~\bibnamefont {Cros}},\ }\bibfield
  {title} {\bibinfo {title} {{Magnetic skyrmions: advances in physics and
  potential applications}},\ }\href
  {https://doi.org/10.1038/natrevmats.2017.31} {\bibfield  {journal} {\bibinfo
  {journal} {Nat. Rev. Mater.}\ }\textbf {\bibinfo {volume} {2}},\ \bibinfo
  {pages} {17031} (\bibinfo {year} {2017})}\BibitemShut {NoStop}%
\bibitem [{\citenamefont {Nagaosa}\ and\ \citenamefont
  {Tokura}(2013)}]{Nagaosa2013}%
  \BibitemOpen
  \bibfield  {author} {\bibinfo {author} {\bibfnamefont {N.}~\bibnamefont
  {Nagaosa}}\ and\ \bibinfo {author} {\bibfnamefont {Y.}~\bibnamefont
  {Tokura}},\ }\bibfield  {title} {\bibinfo {title} {{Topological properties
  and dynamics of magnetic skyrmions}},\ }\href
  {https://doi.org/10.1038/nnano.2013.243} {\bibfield  {journal} {\bibinfo
  {journal} {Nat. Nanotechnol.}\ }\textbf {\bibinfo {volume} {8}},\ \bibinfo
  {pages} {899} (\bibinfo {year} {2013})}\BibitemShut {NoStop}%
\bibitem [{Note1()}]{Note1}%
  \BibitemOpen
  \bibinfo {note} {This convention of meron and antimeron is the same as most
  of the papers on the subject except Ref. \cite {Yu2018}, where a different
  convention is used.}\BibitemShut {Stop}%
\bibitem [{\citenamefont {Kovalev}\ and\ \citenamefont
  {Sandhoefner}(2018)}]{Kovalev2018}%
  \BibitemOpen
  \bibfield  {author} {\bibinfo {author} {\bibfnamefont {A.~A.}\ \bibnamefont
  {Kovalev}}\ and\ \bibinfo {author} {\bibfnamefont {S.}~\bibnamefont
  {Sandhoefner}},\ }\bibfield  {title} {\bibinfo {title} {{Skyrmions and
  Antiskyrmions in Quasi-Two-Dimensional Magnets}},\ }\href
  {https://doi.org/10.3389/fphy.2018.00098} {\bibfield  {journal} {\bibinfo
  {journal} {Front. Phys.}\ }\textbf {\bibinfo {volume} {6}},\ \bibinfo {pages}
  {1} (\bibinfo {year} {2018})}\BibitemShut {NoStop}%
\bibitem [{\citenamefont {Fano}(1949)}]{Fano1949}%
  \BibitemOpen
  \bibfield  {author} {\bibinfo {author} {\bibfnamefont {U.}~\bibnamefont
  {Fano}},\ }\bibfield  {title} {\bibinfo {title} {{Remarks on the Classical
  and Quantum-Mechanical Treatment of Partial Polarization}},\ }\href
  {https://doi.org/10.1364/JOSA.39.000859} {\bibfield  {journal} {\bibinfo
  {journal} {J. Opt. Soc. Am.}\ }\textbf {\bibinfo {volume} {39}},\ \bibinfo
  {pages} {859} (\bibinfo {year} {1949})}\BibitemShut {NoStop}%
\bibitem [{\citenamefont {Falkoff}\ and\ \citenamefont
  {MacDonald}(1951)}]{Falkoff1951}%
  \BibitemOpen
  \bibfield  {author} {\bibinfo {author} {\bibfnamefont {D.~L.}\ \bibnamefont
  {Falkoff}}\ and\ \bibinfo {author} {\bibfnamefont {J.~E.}\ \bibnamefont
  {MacDonald}},\ }\bibfield  {title} {\bibinfo {title} {{On the Stokes
  Parameters for Polarized Radiation}},\ }\href
  {https://doi.org/10.1364/JOSA.41.000861} {\bibfield  {journal} {\bibinfo
  {journal} {J. Opt. Soc. Am.}\ }\textbf {\bibinfo {volume} {41}},\ \bibinfo
  {pages} {861} (\bibinfo {year} {1951})}\BibitemShut {NoStop}%
\bibitem [{\citenamefont {Penrose}(2005)}]{penrose-roadtoreality-2005}%
  \BibitemOpen
  \bibfield  {author} {\bibinfo {author} {\bibfnamefont {R.}~\bibnamefont
  {Penrose}},\ }\href@noop {} {\emph {\bibinfo {title} {{The road to reality :
  a complete guide to the laws of the universe}}}}\ (\bibinfo  {publisher}
  {Vintage},\ \bibinfo {address} {London},\ \bibinfo {year} {2005})\BibitemShut
  {NoStop}%
\bibitem [{\citenamefont {Dong}\ \emph {et~al.}(2017)\citenamefont {Dong},
  \citenamefont {Chen}, \citenamefont {Zhu}, \citenamefont {Wang},\ and\
  \citenamefont {Zhang}}]{Dong2017}%
  \BibitemOpen
  \bibfield  {author} {\bibinfo {author} {\bibfnamefont {J.-W.}\ \bibnamefont
  {Dong}}, \bibinfo {author} {\bibfnamefont {X.-D.}\ \bibnamefont {Chen}},
  \bibinfo {author} {\bibfnamefont {H.}~\bibnamefont {Zhu}}, \bibinfo {author}
  {\bibfnamefont {Y.}~\bibnamefont {Wang}},\ and\ \bibinfo {author}
  {\bibfnamefont {X.}~\bibnamefont {Zhang}},\ }\bibfield  {title} {\bibinfo
  {title} {{Valley photonic crystals for control of spin and topology}},\
  }\href {https://doi.org/10.1038/nmat4807} {\bibfield  {journal} {\bibinfo
  {journal} {Nat. Mater.}\ }\textbf {\bibinfo {volume} {16}},\ \bibinfo {pages}
  {298} (\bibinfo {year} {2017})}\BibitemShut {NoStop}%
\bibitem [{\citenamefont {Chen}\ \emph
  {et~al.}(2017{\natexlab{a}})\citenamefont {Chen}, \citenamefont {Zhao},
  \citenamefont {Chen},\ and\ \citenamefont {Dong}}]{Chen2017a}%
  \BibitemOpen
  \bibfield  {author} {\bibinfo {author} {\bibfnamefont {X.-D.}\ \bibnamefont
  {Chen}}, \bibinfo {author} {\bibfnamefont {F.-l.}\ \bibnamefont {Zhao}},
  \bibinfo {author} {\bibfnamefont {M.}~\bibnamefont {Chen}},\ and\ \bibinfo
  {author} {\bibfnamefont {J.-w.}\ \bibnamefont {Dong}},\ }\bibfield  {title}
  {\bibinfo {title} {{Valley-contrasting physics in all-dielectric photonic
  crystals: Orbital angular momentum and topological propagation}},\ }\href
  {https://doi.org/10.1103/PhysRevB.96.020202} {\bibfield  {journal} {\bibinfo
  {journal} {Phys. Rev. B}\ }\textbf {\bibinfo {volume} {96}},\ \bibinfo
  {pages} {020202} (\bibinfo {year} {2017}{\natexlab{a}})}\BibitemShut
  {NoStop}%
\bibitem [{\citenamefont {Xiao}\ \emph {et~al.}(2007)\citenamefont {Xiao},
  \citenamefont {Yao},\ and\ \citenamefont {Niu}}]{Xiao2007}%
  \BibitemOpen
  \bibfield  {author} {\bibinfo {author} {\bibfnamefont {D.}~\bibnamefont
  {Xiao}}, \bibinfo {author} {\bibfnamefont {W.}~\bibnamefont {Yao}},\ and\
  \bibinfo {author} {\bibfnamefont {Q.}~\bibnamefont {Niu}},\ }\bibfield
  {title} {\bibinfo {title} {{Valley-Contrasting Physics in Graphene: Magnetic
  Moment and Topological Transport}},\ }\href
  {https://doi.org/10.1103/PhysRevLett.99.236809} {\bibfield  {journal}
  {\bibinfo  {journal} {Phys. Rev. Lett.}\ }\textbf {\bibinfo {volume} {99}},\
  \bibinfo {pages} {236809} (\bibinfo {year} {2007})}\BibitemShut {NoStop}%
\bibitem [{\citenamefont {Cao}\ \emph {et~al.}(2012)\citenamefont {Cao},
  \citenamefont {Wang}, \citenamefont {Han}, \citenamefont {Ye}, \citenamefont
  {Zhu}, \citenamefont {Shi}, \citenamefont {Niu}, \citenamefont {Tan},
  \citenamefont {Wang}, \citenamefont {Liu},\ and\ \citenamefont
  {Feng}}]{Cao2012}%
  \BibitemOpen
  \bibfield  {author} {\bibinfo {author} {\bibfnamefont {T.}~\bibnamefont
  {Cao}}, \bibinfo {author} {\bibfnamefont {G.}~\bibnamefont {Wang}}, \bibinfo
  {author} {\bibfnamefont {W.}~\bibnamefont {Han}}, \bibinfo {author}
  {\bibfnamefont {H.}~\bibnamefont {Ye}}, \bibinfo {author} {\bibfnamefont
  {C.}~\bibnamefont {Zhu}}, \bibinfo {author} {\bibfnamefont {J.}~\bibnamefont
  {Shi}}, \bibinfo {author} {\bibfnamefont {Q.}~\bibnamefont {Niu}}, \bibinfo
  {author} {\bibfnamefont {P.}~\bibnamefont {Tan}}, \bibinfo {author}
  {\bibfnamefont {E.}~\bibnamefont {Wang}}, \bibinfo {author} {\bibfnamefont
  {B.}~\bibnamefont {Liu}},\ and\ \bibinfo {author} {\bibfnamefont
  {J.}~\bibnamefont {Feng}},\ }\bibfield  {title} {\bibinfo {title}
  {{Valley-selective circular dichroism of monolayer molybdenum disulphide}},\
  }\href {https://doi.org/10.1038/ncomms1882} {\bibfield  {journal} {\bibinfo
  {journal} {Nat. Commun.}\ }\textbf {\bibinfo {volume} {3}},\ \bibinfo {pages}
  {887} (\bibinfo {year} {2012})}\BibitemShut {NoStop}%
\bibitem [{\citenamefont {Chen}\ \emph
  {et~al.}(2017{\natexlab{b}})\citenamefont {Chen}, \citenamefont {Liu},
  \citenamefont {Zhang}, \citenamefont {Wang}, \citenamefont {Guan},
  \citenamefont {Liu}, \citenamefont {Shi}, \citenamefont {Lu},\ and\
  \citenamefont {Zi}}]{Chen2017}%
  \BibitemOpen
  \bibfield  {author} {\bibinfo {author} {\bibfnamefont {A.}~\bibnamefont
  {Chen}}, \bibinfo {author} {\bibfnamefont {W.}~\bibnamefont {Liu}}, \bibinfo
  {author} {\bibfnamefont {Y.}~\bibnamefont {Zhang}}, \bibinfo {author}
  {\bibfnamefont {B.}~\bibnamefont {Wang}}, \bibinfo {author} {\bibfnamefont
  {F.}~\bibnamefont {Guan}}, \bibinfo {author} {\bibfnamefont {X.}~\bibnamefont
  {Liu}}, \bibinfo {author} {\bibfnamefont {L.}~\bibnamefont {Shi}}, \bibinfo
  {author} {\bibfnamefont {L.}~\bibnamefont {Lu}},\ and\ \bibinfo {author}
  {\bibfnamefont {J.}~\bibnamefont {Zi}},\ }\bibfield  {title} {\bibinfo
  {title} {{Observing half and integer polarization vortices at band
  degeneracies}},\ }\href {http://arxiv.org/abs/1712.09296} {\  (\bibinfo
  {year} {2017}{\natexlab{b}})},\ \Eprint {https://arxiv.org/abs/1712.09296}
  {arXiv:1712.09296} \BibitemShut {NoStop}%
\bibitem [{\citenamefont {Ma}\ and\ \citenamefont {Shvets}(2016)}]{Ma2016}%
  \BibitemOpen
  \bibfield  {author} {\bibinfo {author} {\bibfnamefont {T.}~\bibnamefont
  {Ma}}\ and\ \bibinfo {author} {\bibfnamefont {G.}~\bibnamefont {Shvets}},\
  }\bibfield  {title} {\bibinfo {title} {{All-Si valley-Hall photonic
  topological insulator}},\ }\href
  {https://doi.org/10.1088/1367-2630/18/2/025012} {\bibfield  {journal}
  {\bibinfo  {journal} {New J. Phys.}\ }\textbf {\bibinfo {volume} {18}},\
  \bibinfo {pages} {025012} (\bibinfo {year} {2016})}\BibitemShut {NoStop}%
\bibitem [{\citenamefont {Khanikaev}\ \emph {et~al.}(2013)\citenamefont
  {Khanikaev}, \citenamefont {{Hossein Mousavi}}, \citenamefont {Tse},
  \citenamefont {Kargarian}, \citenamefont {MacDonald},\ and\ \citenamefont
  {Shvets}}]{Khanikaev2013}%
  \BibitemOpen
  \bibfield  {author} {\bibinfo {author} {\bibfnamefont {A.~B.}\ \bibnamefont
  {Khanikaev}}, \bibinfo {author} {\bibfnamefont {S.}~\bibnamefont {{Hossein
  Mousavi}}}, \bibinfo {author} {\bibfnamefont {W.-k.}\ \bibnamefont {Tse}},
  \bibinfo {author} {\bibfnamefont {M.}~\bibnamefont {Kargarian}}, \bibinfo
  {author} {\bibfnamefont {A.~H.}\ \bibnamefont {MacDonald}},\ and\ \bibinfo
  {author} {\bibfnamefont {G.}~\bibnamefont {Shvets}},\ }\bibfield  {title}
  {\bibinfo {title} {{Photonic topological insulators}},\ }\href
  {https://doi.org/10.1038/nmat3520} {\bibfield  {journal} {\bibinfo  {journal}
  {Nat. Mater.}\ }\textbf {\bibinfo {volume} {12}},\ \bibinfo {pages} {233}
  (\bibinfo {year} {2013})}\BibitemShut {NoStop}%
\bibitem [{\citenamefont {Fan}\ and\ \citenamefont
  {Joannopoulos}(2002)}]{Fan2002}%
  \BibitemOpen
  \bibfield  {author} {\bibinfo {author} {\bibfnamefont {S.}~\bibnamefont
  {Fan}}\ and\ \bibinfo {author} {\bibfnamefont {J.~D.}\ \bibnamefont
  {Joannopoulos}},\ }\bibfield  {title} {\bibinfo {title} {{Analysis of guided
  resonances in photonic crystal slabs}},\ }\href
  {https://doi.org/10.1103/PhysRevB.65.235112} {\bibfield  {journal} {\bibinfo
  {journal} {Phys. Rev. B}\ }\textbf {\bibinfo {volume} {65}},\ \bibinfo
  {pages} {235112} (\bibinfo {year} {2002})}\BibitemShut {NoStop}%
\bibitem [{\citenamefont {Wimmer}\ \emph {et~al.}(2017)\citenamefont {Wimmer},
  \citenamefont {Price}, \citenamefont {Carusotto},\ and\ \citenamefont
  {Peschel}}]{wimmer2017experimental}%
  \BibitemOpen
  \bibfield  {author} {\bibinfo {author} {\bibfnamefont {M.}~\bibnamefont
  {Wimmer}}, \bibinfo {author} {\bibfnamefont {H.~M.}\ \bibnamefont {Price}},
  \bibinfo {author} {\bibfnamefont {I.}~\bibnamefont {Carusotto}},\ and\
  \bibinfo {author} {\bibfnamefont {U.}~\bibnamefont {Peschel}},\ }\bibfield
  {title} {\bibinfo {title} {Experimental measurement of the berry curvature
  from anomalous transport},\ }\href@noop {} {\bibfield  {journal} {\bibinfo
  {journal} {Nature Physics}\ }\textbf {\bibinfo {volume} {13}},\ \bibinfo
  {pages} {545} (\bibinfo {year} {2017})}\BibitemShut {NoStop}%
\bibitem [{\citenamefont {Guo}\ \emph {et~al.}(2019)\citenamefont {Guo},
  \citenamefont {Xiao}, \citenamefont {Zhou},\ and\ \citenamefont
  {Fan}}]{Guo2019}%
  \BibitemOpen
  \bibfield  {author} {\bibinfo {author} {\bibfnamefont {Y.}~\bibnamefont
  {Guo}}, \bibinfo {author} {\bibfnamefont {M.}~\bibnamefont {Xiao}}, \bibinfo
  {author} {\bibfnamefont {Y.}~\bibnamefont {Zhou}},\ and\ \bibinfo {author}
  {\bibfnamefont {S.}~\bibnamefont {Fan}},\ }\bibfield  {title} {\bibinfo
  {title} {{Arbitrary Polarization Conversion with a Photonic Crystal Slab}},\
  }\href {https://doi.org/10.1002/adom.201801453} {\bibfield  {journal}
  {\bibinfo  {journal} {Adv. Opt. Mater.}\ }\textbf {\bibinfo {volume}
  {1801453}},\ \bibinfo {pages} {1801453} (\bibinfo {year} {2019})}\BibitemShut
  {NoStop}%
\bibitem [{\citenamefont {Zhang}\ \emph {et~al.}(2018)\citenamefont {Zhang},
  \citenamefont {Chen}, \citenamefont {Liu}, \citenamefont {Hsu}, \citenamefont
  {Wang}, \citenamefont {Guan}, \citenamefont {Liu}, \citenamefont {Shi},
  \citenamefont {Lu},\ and\ \citenamefont {Zi}}]{Zhang2018}%
  \BibitemOpen
  \bibfield  {author} {\bibinfo {author} {\bibfnamefont {Y.}~\bibnamefont
  {Zhang}}, \bibinfo {author} {\bibfnamefont {A.}~\bibnamefont {Chen}},
  \bibinfo {author} {\bibfnamefont {W.}~\bibnamefont {Liu}}, \bibinfo {author}
  {\bibfnamefont {C.~W.}\ \bibnamefont {Hsu}}, \bibinfo {author} {\bibfnamefont
  {B.}~\bibnamefont {Wang}}, \bibinfo {author} {\bibfnamefont {F.}~\bibnamefont
  {Guan}}, \bibinfo {author} {\bibfnamefont {X.}~\bibnamefont {Liu}}, \bibinfo
  {author} {\bibfnamefont {L.}~\bibnamefont {Shi}}, \bibinfo {author}
  {\bibfnamefont {L.}~\bibnamefont {Lu}},\ and\ \bibinfo {author}
  {\bibfnamefont {J.}~\bibnamefont {Zi}},\ }\bibfield  {title} {\bibinfo
  {title} {{Observation of Polarization Vortices in Momentum Space}},\ }\href
  {https://doi.org/10.1103/PhysRevLett.120.186103} {\bibfield  {journal}
  {\bibinfo  {journal} {Phys. Rev. Lett.}\ }\textbf {\bibinfo {volume} {120}},\
  \bibinfo {pages} {186103} (\bibinfo {year} {2018})}\BibitemShut {NoStop}%
\bibitem [{\citenamefont {Doeleman}\ \emph {et~al.}(2018)\citenamefont
  {Doeleman}, \citenamefont {Monticone}, \citenamefont {den Hollander},
  \citenamefont {Al{\`{u}}},\ and\ \citenamefont {Koenderink}}]{Doeleman2018}%
  \BibitemOpen
  \bibfield  {author} {\bibinfo {author} {\bibfnamefont {H.~M.}\ \bibnamefont
  {Doeleman}}, \bibinfo {author} {\bibfnamefont {F.}~\bibnamefont {Monticone}},
  \bibinfo {author} {\bibfnamefont {W.}~\bibnamefont {den Hollander}}, \bibinfo
  {author} {\bibfnamefont {A.}~\bibnamefont {Al{\`{u}}}},\ and\ \bibinfo
  {author} {\bibfnamefont {A.~F.}\ \bibnamefont {Koenderink}},\ }\bibfield
  {title} {\bibinfo {title} {{Experimental observation of a polarization vortex
  at an optical bound state in the continuum}},\ }\href
  {https://doi.org/10.1038/s41566-018-0177-5} {\bibfield  {journal} {\bibinfo
  {journal} {Nat. Photonics}\ }\textbf {\bibinfo {volume} {12}},\ \bibinfo
  {pages} {397} (\bibinfo {year} {2018})}\BibitemShut {NoStop}%
\end{thebibliography}
\end{document}